\documentclass[aps,preprint,tightenlines,showpacs,nofootinbib]{revtex4}

\usepackage{graphicx}  
\usepackage{bm}  
\usepackage{amsmath}
\newcommand{\beq}{\begin{equation}}
\newcommand{\eeq}{\end{equation}}
\newcommand{\bea}{\begin{eqnarray}}
\newcommand{\eea}{\end{eqnarray}}       

\newcommand{\simlt}{\stackrel{<}{{}_\sim}}

\def\bpp{{b_1^{1/2,\,1/2}}}
\def\bpm{{b_1^{1/2,\,-1/2}}}
\def\bppm{{b_1^{1/2,\,\pm 1/2}}}

\begin{document}
\title{Dispersion analysis of the nucleon form factors\\
including meson continua}
\author{M.A. Belushkin}\email{belushki@itkp.uni-bonn.de}
\author{H.-W. Hammer}\email{hammer@itkp.uni-bonn.de}
\affiliation{Helmholtz-Institut f\"ur Strahlen- und Kernphysik (Theorie),
Universit\"at Bonn, Nu\ss allee 14-16,\\ D-53115 Bonn, Germany}

\author{Ulf-G. Mei\ss ner}\email{meissner@itkp.uni-bonn.de}
\affiliation{Helmholtz-Institut f\"ur Strahlen- und Kernphysik (Theorie),
Universit\"at Bonn, Nu\ss allee 14-16, D-53115 Bonn, Germany\\
and\\
Institut f\"ur Kernphysik (Theorie), Forschungszentrum
J\"ulich, D-52425 J\"ulich, Germany\\}

\date{\today}
\begin{abstract}
Dispersion relations provide a powerful tool to analyse the
electromagnetic form factors of the nucleon for all momentum transfers.
Constraints from meson-nucleon scattering data, unitarity, 
and perturbative QCD can be included in a straightforward way.
In particular, we include  the $2\pi$, $\rho\pi$, and $K\bar{K}$ continua
as independent input in our analysis and provide an error band for
our results.
Moreover, we discuss two different methods to include the asymptotic
constraints from perturbative QCD.
We simultaneously analyze the world data for all four form factors
in both the space-like and time-like regions and generally find 
good agreement with the data. We also extract
the nucleon radii and the $\omega NN$ coupling constants.
For the radii, we generally find
good agreement with other determinations with the exception
of the electric charge radius of the proton which comes out smaller. 
The $\omega NN$ vector coupling constant is determined 
relatively well by the fits, but for the tensor coupling constant even 
the sign can not be determined.
\end{abstract}
\pacs{11.55.Fv, 13.40.Gp, 14.20.Dh}
\maketitle
\section{Introduction}
\label{sec:intro}

The electromagnetic (em) form factors of the nucleon provide an important tool
to study strong interaction dynamics over a wide range of momentum
transfers \cite{Gao:2003ag,Hyde-Wright:2004gh}.
Their detailed understanding is important for unraveling aspects of 
perturbative and nonperturbative nucleon structure. 
At small momentum transfers, they are determined by
gross properties of the nucleon like the charge and magnetic moment.
At high momentum transfers they encode information on
the quark substructure of the nucleon as described by perturbative QCD.
The form factors also contain
important information on nucleon radii and vector meson coupling
constants. Moreover, they are an
important ingredient in a wide range of experiments from
Lamb shift measurements \cite{Ude97} to determinations of the
strangeness content of the nucleon \cite{strange}.

With the advent of the new continuous beam electron accelerators
such as CEBAF (Jefferson Laboratory), ELSA (Bonn), and MAMI (Mainz), a
wealth of precise data for space-like momentum transfers
has become available \cite{Ostrick}. 
Due to the difficulty of the experiments, the
time-like form factors are less well known. While there is a fair 
amount of information on the proton time-like form factors 
\cite{Ambrogiani:1999bh,Baldini},
only one measurement of the neutron form factor from the pioneering 
FENICE experiment \cite{Antonelli:1993vz,Antonelli:1998fv} exists. 
Recently, new precise data on the proton time-like form factors have been
presented by the BES, CLEO, and BABAR collaborations
\cite{Ablikim:2005,Pedlar:2005sj,Aubert:2005cb}.

In this work, we therefore analyze the nucleon form factors in both
the space- and time-like regions. We use dispersion 
relations which provide a model-independent framework to consistently 
analyze the form factor data in both regions.
A complete description of our data base is given in section \ref{sec:fits}.

It has been known for a long time that the pion plays an important role in 
the long-range structure of the nucleon \cite{FHK}. This connection
was made more precise using dispersion theory in the 1950's
\cite{CKGZ58,FGT58}. Subsequently, Frazer and Fulco have written down 
partial wave dispersion relations that relate the nucleon electromagnetic 
structure to pion-nucleon ($\pi N$) scattering and predicted the existence of
the $\rho$ resonance \cite{FF,FF60a}.
H\"ohler et al. \cite{Hohler:1976ax} first performed a 
dispersion analysis of the electromagnetic form factors of the
nucleon including the $2\pi$ continuum
derived from the pion form factor and $\pi N$-scattering data \cite{HP}.
In the mid-nineties, this analysis has been updated 
by Mergell, Mei\ss ner, and Drechsel \cite{MMD96} and was later
extended to include data in the time-like region \cite{HMD96,slacproc}. 
The new precise data for the neutron electric form factor have 
been included as well \cite{HM04}.
For recent attempts to calculate the nucleon form factors
in QCD using nonperturbative methods, see e.g. 
Refs.~\cite{Braun06,Gockeler:2003ay,Negele06}.

A recent form factor analysis by Friedrich and Walcher \cite{FW}
created some renewed interest in the $2\pi$ continuum. They 
analysed the electromagnetic nucleon form factors and performed
various phenomenological fits \cite{FW}. Their fits showed a pronounced
bump-dip structure in $G_E^n$ which they interpreted as a signature 
of a very long-range contribution of the pion cloud to the charge distribution 
in the Breit frame extending out to about 2~fm.
This observation is at variance with the pion cloud contribution to the 
nucleon form factors as given by the $2\pi$ continuum
-- the lowest-mass intermediate state including pions only
\cite{Hammer:2003qv}. We will discuss a possibility how to
reconcile these findings below.

It is well known that vector mesons play an important 
role in the electromagnetic structure of the nucleon, see e.g. 
Refs.~\cite{FF,Sak,GoSa,Gari,Lomo,Dubni,Tomasi-Gustafsson:2005kc}, 
and the remaining contributions to the spectral function
have usually been approximated by vector meson resonances. 
A novel addition in this work is the inclusion of the $K\bar{K}$ 
\cite{Hammer:1998rz,Hammer:1999uf} and $\rho\pi$ \cite{Meissner:1997qt}
continuum contributions similar to the $2\pi$ continuum described above.
The continuum contributions provide independent information on the
spectral functions that supplements the electromagnetic form factor data.
Moreover, we enforce the asymptotic constraints from pQCD.
We discuss two different approaches to obtain this
behavior: superconvergence relations
and an explicit continuum term with the correct pQCD behavior, and
show results for both methods.

Our manuscript is organized as follows. Sec.~\ref{sec:def} contains the basic 
definitions of the nucleon em form factors
and a short discussion of the corresponding dispersion relations. The
various continuum contributions to the spectral functions are discussed in 
Sec.~\ref{sec:2pico}. The structure of and the constraints on the spectral
functions are given in Sec.~\ref{sec:spec}. The results of our fits are
presented and discussed in Sec.\ref{sec:fits}. We close with a short
summary and outlook in Sec.~\ref{sec:summ}. The fit parameters are collected
in the appendix.

\section{Preliminaries}

\subsection{Definitions}
\label{sec:def}

\begin{figure}[ht] 
\centerline{\includegraphics*[width=4.5cm,angle=0]{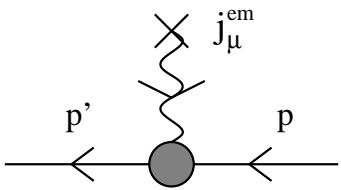}}
\caption{
The nucleon matrix element of the electromagnetic current $j_\mu^{\rm em}$.}
\label{fig:curr}
\end{figure}

The electromagnetic (em) structure of the nucleon is determined by
the matrix element of the vector current operator $j_\mu^{\rm em}$ 
between nucleon states as illustrated in Fig.~\ref{fig:curr}.

Using Lorentz and gauge invariance, this matrix element can  
be expressed in terms of two form factors,
\begin{equation}
\langle p' | j_\mu^{\rm em} | p \rangle = \bar{u}(p')
\left[ F_1 (t) \gamma_\mu +i\frac{F_2 (t)}{2 M} \sigma_{\mu\nu}
q^\nu \right] u(p)\,,
\end{equation}
where $M$ is the nucleon mass and $t=(p'-p)^2$ 
the four-momentum transfer squared. For data in the space-like region, it is
often convenient to use the variable $Q^2=-t>0$.
The functions
$F_1(t)$ and $F_2(t)$ are the Dirac and Pauli form factors, respectively.
They are normalized at $t=0$ as
\begin{equation}
\label{norm}
F_1^p(0) = 1\,, \; F_1^n(0) = 0\,, \; F_2^p(0) =  \kappa_p\,,
\; F_2^n(0) = \kappa_n\, ,
\end{equation}
with $\kappa_p=1.793$ and $\kappa_n=-1.913$ the anomalous magnetic moment of
the proton and the neutron, respectively, in units of nuclear magnetons.

It is convenient to work in the isospin basis and to 
decompose the form factors into isoscalar and isovector parts,
\begin{equation}
F_i^s = \frac{1}{2} (F_i^p + F_i^n) \, , \quad
F_i^v = \frac{1}{2} (F_i^p - F_i^n) \, ,
\end{equation}
where $i = 1,2 \,$. 
The experimental data are usually given for the Sachs form factors
\begin{eqnarray}
G_{E}(t) &=& F_1(t) - \tau F_2(t) \, , \nonumber\\
G_{M}(t) &=& F_1(t) + F_2(t) \, , 
\label{sachs}
\end{eqnarray}
where $\tau = -t/(4 M^2)$.
In the Breit frame, $G_{E}$ and $G_{M}$ may be interpreted as
the Fourier transforms of the charge and magnetization distributions,
respectively.              

The nucleon radii $\sqrt{\langle r^2 \rangle}$
can be defined from the low-$t$ expansion
of the form factors,
\beq
F(t)=F(0)\left[1+t\langle r^2 \rangle /6 +\ldots \right]\,,
\eeq
where $F(t)$ is a generic form factor. In the case of the electric
and Dirac form factors of the neutron, $G_E^n$ and $F_1^n$, the
expansion starts with the term linear in $t$ and the 
normalization factor $F(0)$ is dropped.

\subsection{Dispersion Relations and Spectral Decomposition}
\label{sec:specdeco}

Based on unitarity and analyticity, dispersion relations relate
the real and imaginary parts of the electromagnetic nucleon form factors. 
Let $F(t)$ be a generic symbol for any one of the four independent 
nucleon form factors. We write down an unsubtracted
dispersion relation of the form
\begin{equation}
F(t) = \frac{1}{\pi} \, \int_{t_0}^\infty \frac{{\rm Im}\, 
F(t')}{t'-t-i\epsilon}\, dt'\, ,
\label{emff:disp} 
\end{equation}
where $t_0$ is the threshold of the lowest cut of $F(t)$ (see below)
and the $i\epsilon$ defines the integral for values of $t$ on the 
cut. The convergence of an unsubtracted dispersion relation
for the form factors has been assumed. We could also use 
a once subtracted dispersion relation,
since the normalization of the form factors at $t=0$ is known.
Using Eq.~(\ref{emff:disp}) the electromagnetic structure
of the nucleon can be related to its absorptive behavior.

The imaginary part ${\rm Im}\, F$ entering Eq.~(\ref{emff:disp}) 
can be obtained from a spectral decomposition \cite{CKGZ58,FGT58}. 
For this purpose it is most convenient to consider the 
electromagnetic 
current matrix element in the time-like region ($t>0$), which is 
related to the space-like region ($t<0$) via crossing symmetry.
The matrix element can be expressed as
\begin{eqnarray}
\label{eqJ}
J_\mu &=& \langle N(p) \overline{N}(\bar{p}) | j_\mu^{\rm em}(0) | 0 \rangle \\
&=& \bar{u}(p) \left[ F_1 (t) \gamma_\mu +i\frac{F_2 (t)}{2 M} \sigma_{\mu\nu}
(p+\bar{p})^\nu \right] v(\bar{p})\,,\nonumber
\end{eqnarray}
where $p$ and $\bar{p}$ are the momenta of the nucleon and an\-ti\-nuc\-le\-on
created by the current $j_\mu^{\rm em}$, respectively. 
The four-momentum transfer squared
in the time-like region is $t=(p+\bar{p})^2$. 

\begin{figure}[htb] 
\centerline{\includegraphics*[width=8cm,angle=0]{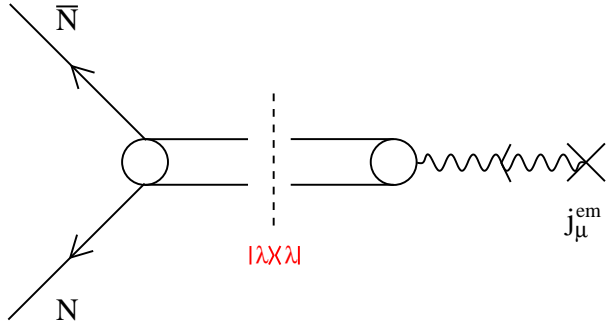}}
\caption{
The spectral decomposition of the 
nucleon matrix element of the electromagnetic current $j_\mu^{\rm em}$.
$| \lambda \rangle $ denotes an hadronic intermediate state.
}
\label{fig:spec}
\end{figure}

Using the LSZ reduction formalism, the imaginary part
of the form factors is obtained by inserting a complete set of
intermediate states as \cite{CKGZ58,FGT58}
\begin{eqnarray}
\label{spectro}
{\rm Im}\,J_\mu &=& \frac{\pi}{Z}(2\pi)^{3/2}{\cal N}\,\sum_\lambda
 \langle p | \bar{J}_N (0) | \lambda \rangle 
\langle \lambda | j_\mu^{\rm em} (0) | 0 \rangle \,v(\bar{p})
\,\delta^4(p+\bar{p}-p_\lambda)\,,
\end{eqnarray}
where ${\cal N}$ is a nucleon spinor normalization factor, $Z$ is
the nucleon wave function renormalization, and $\bar{J}_N (x) =
J^\dagger(x) \gamma_0$ with $J_N(x)$ a nucleon source.
This decomposition is illustrated in Fig.~\ref{fig:spec}.
It relates the spectral function to on-shell matrix elements of other
processes.

The states $|\lambda\rangle$ are asymptotic states of
momentum $p_\lambda$. They carry the same quantum numbers as
the current $j^{\rm em}_\mu$: $I^G(J^{PC})=0^-(1^{--})$ for
the isoscalar current and $I^G(J^{PC})=1^+(1^{--})$ for the
isovector component of $j^{\rm em}_\mu$. 
Furthermore, they have zero net baryon number. Because of $G$-parity, states
with an odd number of pions only contribute to the iso\-scalar
part, while states with an even number contribute to the 
isovector part.
For the isoscalar part  the lowest mass states are: $3\pi$,
$5\pi$, $\ldots$, $K\bar{K}$, $K\bar{K}\pi$, $\ldots$; 
for the isovector part they are: $2\pi$, $4\pi$, $\ldots$. 

Associated with each intermediate state is a
cut starting at the corresponding threshold in $t$ and running to
infinity. As a consequence,
the spectral function ${\rm Im}\, F(t)$ is different from zero along the
cut from $t_0$ to $\infty$, with $t_0 = 4 \, (9) \, M_\pi^2$ for the
isovector (isoscalar) case.

The spectral functions are the central quantities in the 
dispersion-theoretical approach. Using Eqs.~(\ref{eqJ},\ref{spectro}), they
can in principle be constructed from experimental data. 
In practice, this program can only be carried out for 
the lightest two-particle intermediate states.

The longest-range, and therefore at low momentum transfer most 
important continuum contribution comes from the $2\pi$ intermediate state
which contributes to the isovector form factors \cite{HP}.
A new calculation of this contribution 
has recently been performed in Ref.~\cite{Belushkin:2005ds}.
In this analysis we for the first time also include the $K\bar{K}$ and 
$\rho\pi$ continua 
\cite{Hammer:1998rz,Hammer:1999uf,Meissner:1997qt}.\footnote{Note
that the effect of these continua has previously been studied in
Ref.~\cite{Hammer:1999uf} using fits to parameterizations of form 
factor data.}

\section{Continuum Contributions}

Our general strategy is to include as much physics information in the 
construction of the spectral functions as possible.
In this section, we explicitly contruct the $2\pi$, $K\bar{K}$, and 
$\rho\pi$ continua mentioned above. These continua will be an important part
of our spectral functions.

\subsection{$2\pi$ Continuum}
\label{sec:2pico}

The $2\pi$ contribution has recently been reevaluated in a 
model--independent way \cite{Belushkin:2005ds}
using the latest experimental data for the pion
form factor from CMD-2 \cite{CMD2}, KLOE \cite{KLOE}, and
SND \cite{SND}. Here we give a short summary of this evaluation.

Following  Refs.~\cite{LB,Belushkin:2005ds},
the $2\pi$ contribution to the
isovector spectral functions  in terms of the pion 
charge form factor $F_\pi (t)$ and the P--wave $\pi\pi \to \bar N N$ 
amplitudes $f^1_\pm(t)$ can be expressed as:
\bea 
\nonumber 
{\rm Im}~G_E^{v} (t) &=& \frac{q_t^3}{M\sqrt{t}}\, 
F_\pi (t)^* \, f^1_+ (t)~,\\ 
{\rm Im}~G_M^{v} (t) &=& \frac{q_t^3}{\sqrt{2t}}\, 
F_\pi (t)^* \, f^1_- (t)~, 
\label{uni} 
\eea 
where $q_t=\sqrt{t/4-M_\pi^2}$. The imaginary parts of the Dirac
and Pauli Form factors can be obtained using Eq.~(\ref{sachs}).
The $2\pi$ continuum is expected to be the
dominant contribution to the isovector spectral function 
from threshold up to masses of about $\sqrt{t}\approx 1$~GeV \cite{LB}.

The  P--wave $\pi\pi \to \bar N N$ amplitudes $f_\pm^1(t)$ are tabulated in  
Ref.~\cite{LB}. The representation of Eq.~(\ref{uni}) 
gives the exact isovector spectral functions for $4M_\pi^2 \leq t 
\leq 16 M_\pi^2$, but in practice holds up to $t \simeq 50 M_\pi^2
\approx 1$~GeV$^2$.
Since the contributions from $4\pi$ and higher 
intermediate states is small up to  $t \simeq 50 M_\pi^2$, $F_\pi(t)$
and the $f_\pm^1(t)$ share the same phase in this region and the 
two quantities can be replaced by their absolute values.

The experimental data for the pion
form factor from CMD-2 \cite{CMD2}, KLOE \cite{KLOE}, and
SND \cite{SND} show some discrepancies. In Ref.~\cite{Belushkin:2005ds},
the  $2\pi$ continuum given by Eq.~(\ref{uni}) was evaluated
for all three sets and the errors from the discrepancy 
between the sets were estimated.
The resulting difference in the spectral functions is very small
($\simlt 1\%$). It is largest in the $\rho$-peak region,
but this region is suppressed by the $\pi\pi \to \bar N N$ amplitudes 
$f_\pm^1(t)$ which show a strong fall-off as $t$ increases.

The spectral functions have two distinct features. First, as already 
pointed out in \cite{FF}, they contain the important contribution of 
the $\rho$-meson with its peak at $t \simeq 30 M_\pi^2$. 
Second, on the left shoulder of the $\rho$, the isovector spectral functions 
display a very pronounced enhancement close to the two-pion threshold. This 
is due to the logarithmic singularity on the second Riemann sheet located at 
$t_c = 4M_\pi^2 - M_\pi^4/M^2 = 3.98 M_\pi^2$, very close to the threshold. 
If one were to neglect this important unitarity correction, one would 
severely  underestimate the nucleon isovector radii \cite{HP2}.
In fact, precisely the 
same effect is obtained at leading one-loop accuracy  in 
relativistic chiral perturbation 
theory \cite{GSS,UGMlec}. This topic was also discussed in 
heavy baryon chiral perturbation theory
(ChPT) \cite{BKMspec,Norbert} and in a covariant 
calculation based on infrared regularization \cite{KM,Schindler:2005ke}. 
Thus, the most 
important $2\pi$ contribution to the nucleon form factors can be determined 
by using either unitarity or ChPT (in the latter case, of course, the $\rho$ 
contribution is not included).

The contribution to the nucleon form factors is obtained by inserting
the $2\pi$ contribution to the spectral function into the dispersion relations
Eq.~(\ref{emff:disp}). The result can be parameterized as
\beq
F_i^{(v,2\pi)} (t) = 
\frac{a_i +b_i (1-t/c_i)^{-2/i}}{2(1-t/d_i)}\,, \quad i=1,2\,,
\label{eq:2pipara}
\eeq
where
$a_1=1.10788$, $b_1=0.109364$, $c_1=0.36963$ GeV$^2$, $d_1=0.553034$ GeV$^2$,
$a_2=5.724253$, $b_2=1.111128$, $c_2=0.27175$ GeV$^2$,
and $d_2=0.611258$ GeV$^2$.
The errors in these constants are of the order 4\% or less.
Of course, it is not necessary to 
use a parameterization like Eq.~(\ref{eq:2pipara}) but the numerical
evaluation of the fits becomes much simpler.
The form factor contributions from Eq.~(\ref{eq:2pipara}) are shown
in Fig.~\ref{specff} below.

\subsection{$K\bar{K}$ Continuum}
\label{sec:KKbar}

The $K\bar{K}$ contribution to the isoscalar spectral function was
evaluated in Refs.~\cite{HP,Hammer:1998rz,Hammer:1999uf}
from an analytic continuation of $KN$ scattering data.
In the following, we give a short summary of this work.

The $K\bar{K}$ contribution to the
imaginary part of the isocalar form factors is given by 
\cite{Hammer:1998rz,Hammer:1999uf}
\begin{eqnarray}
\label{imf1}
{\rm Im}\, 
F_1^{(s,K\bar{K})}(t)&=&{\rm Re}\,\left\{\left({M q_t\over 4 p_t^2}\right)
\left[{\sqrt{t}\over 2\sqrt{2}M}\bpm(t)-\bpp(t)\right]
F_K (t)^{\ast}\right\}\,,\\
&& \nonumber \\
\label{imf2}
{\rm Im}\, 
F_2^{(s,K\bar{K})}(t) &=&{\rm Re}\,
\left\{\left({M q_t\over 4 p_t^2}\right)\left[
\bpp(t)-{\sqrt{2}M\over\sqrt{t}}\bpm(t)\right]F_K (t)^{\ast}\right\}\,,
\end{eqnarray}
with
$p_t=\sqrt{t/4-M^2}$ and 
$q_t=\sqrt{t/4-M_K^2}$.
$F_K (t)$ represents the kaon form factor 
whereas the $\bppm (t)$ are the $J=1$ partial wave amplitudes for 
$K\bar{K}\to N\bar{N}$ \cite{Hammer:1998rz,Hammer:1999uf}. Once these imaginary
parts are determined, the contribution of the $K\bar{K}$-continuum 
to the form factors is obtained from the dispersion relation
Eq.~(\ref{emff:disp}). 

For $t \geq 4M^2$ the partial waves are bounded by unitarity,
\begin{equation}
\label{ubs}
\sqrt{p_t / q_t}\,  |\bppm(t)|\leq 1 \,.
\end{equation}
In the unphysical region $4M_K^2 \leq t \leq 4M^2$, however,  they are
not constrained by unitarity.
In Ref.~\cite{Hammer:1998rz}, the amplitudes $\bppm (t)$ 
in the unphysical region
have been determined from an analytic continuation of $KN$-scattering 
amplitudes.
The contribution of the physical region $t\geq  4M^2$ in the dispersion
integral (\ref{emff:disp}) is suppressed for small momentum transfers
and bounded because of Eq.~(\ref{ubs}).
Using the analytically continued amplitudes in the unphysical region
and the unitarity bound in the physical region, the contribution of
the $K\bar{K}$ continuum can therefore be calculated.
Strictly speaking  this calculation provides an upper bound on the 
spectral function since we replace the amplitudes and the 
form factor in Eqs.~(\ref{imf1}, \ref{imf2}) by their absolute
values.

The striking feature in the spectral function is a clear 
$\phi$ resonance structure just above the $K\bar{K}$ threshold.
The resonance structure appears in the partial wave amplitude
$b_1^{1/2,\,1/2}$ as well as in the kaon form factor $F_K$. 
In contrast to the $2\pi$ continuum, there is no strong enhancement on the 
left wing of the $\phi$ resonance which sits directly at the 
$K\bar{K}$ threshold.

The resulting contribution to the nucleon form factors can be 
parameterized by a pole term at the $\phi$ mass:
\beq
F_i^{(s,K\bar{K})}(t) =
\frac{1}{\pi} \int_{4M_K^2}^\infty
\frac{{\rm Im} \, F_i^{(s,K\bar{K})}(t')}{t'-t}dt' \approx
\frac{a_i^{K\bar{K}}}{M_\phi^2-t}\,,
\qquad i=1,2\,,
\label{KKcont}
\eeq
with $a_1^{K\bar{K}}=0.1054$ GeV$^2$ and $a_2^{K\bar{K}}=0.2284$
GeV$^2$.
As a consequence, the contribution of the $K\bar{K}$ continuum 
to the electromagnetic nucleon form factors can conveniently 
be included in the analysis via Eq.~(\ref{KKcont}).
The form factor contributions from Eq.~(\ref{KKcont}) are also shown
in Fig.~\ref{specff} below.

\subsection{$\rho\pi$ Continuum}
\label{sec:rhopi}

Drawing upon a realistic treatment of the correlated $\rho\pi$ 
exchange in the Bonn-J\"ulich $NN$ model \cite{Hol96}, the
$\rho\pi$ contribution to the isoscalar spectral function was
calculated in Ref.~\cite{Meissner:1997qt}. 
The contribution of the $\rho\pi$ continuum can be evaluated
in terms of a dispersion integral which in turn can be represented
by an effective pole term for a fictitious $\omega'$ meson with 
a mass $M_{\omega'} = 1.12$ GeV \cite{Meissner:1997qt}:
\beq
F_i^{(s,\rho\pi)}(t)=
\frac{1}{\pi} \int_{(M_\pi +M_\rho)^2}^\infty
\frac{{\rm Im} \, F_i^{(s,\rho\pi)}(t')}{t'-t}dt' \approx 
\frac{a_i^{\rho\pi}}{M_{\omega'}^2-t}\,, \qquad i=1,2
\eeq
with $a_1^{\rho\pi}=-1.01$ GeV$^2$ and $a_2^{\rho\pi}=-0.04$ GeV$^2$.
In our form factor analysis, we use this effective pole instead of the full 
spectral function.

There is very little sensitivity
in our fits to $a_{2}^{\rho\pi}$, which can vary between $-0.04$ and
$-0.4$ without affecting the outcome of the fit. 
If the $\omega^{\prime}$ pole is treated as a real resonance,
the latter value is consistent with $f_{\omega^{\prime}}\sim 10$ for
$a_{1}^{\rho\pi}=-1.01$ if the coupling constants $g^i_{\omega^{\prime}NN}$
($i=1,2$) from Ref.~\cite{Meissner:1997qt} are used as input.

\begin{figure}[htb]
\centerline{\includegraphics*[width=13.cm,angle=0]{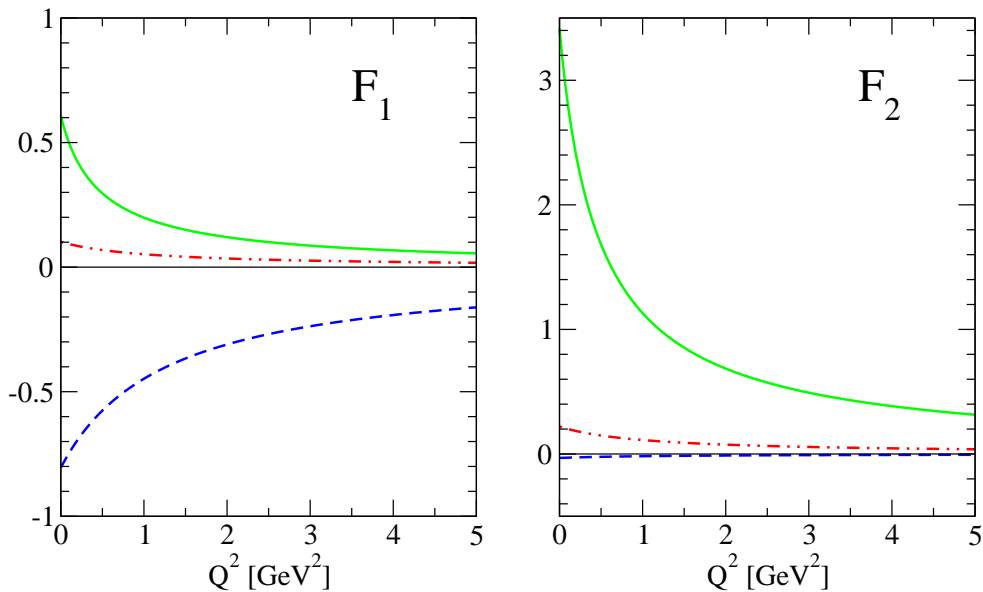}}
\caption{\label{specff} The continuum contributions to the 
nucleon form factors $F_1$ (left panel) and $F_2$ (right panel)
in the space-like region.
The contribution of the $2\pi$ continuum to the isovector form factors 
is given by the solid line, while the contribution of the  
$K\bar{K}$ and $\rho\pi$ continua to the isoscalar form factors are
given by the dash-dotted and dashed lines, respectively.
}
\end{figure}

In Fig.~\ref{specff}, we show the contribution of the 
$2\pi$, $K\bar{K}$, and $\rho\pi$ continua to the electromagnetic 
nucleon form
factors $F_1$ and $F_2$. The  $2\pi$ contributes to the isovector
form factors while the $K\bar{K}$ and $\rho\pi$ continua contribute
to the isoscalar form factors. The $K\bar{K}$ and $\rho\pi$
contributions have opposite sign and partially cancel each other.
The dominant contribution to $F_1^s$ comes from the $\rho\pi$ continuum
while for $F_2^s$ the $K\bar{K}$ contribution is larger.
While the $K\bar{K}$ and $\rho\pi$ contributions can be represented
by simple pole terms, the expressions for the $2\pi$ continuum
Eq.~(\ref{eq:2pipara}) are somewhat more complicated. This is related to the 
strong enhancement close to the $2\pi$ threshold on the left wing of the
$\rho$ resonance discussed above.
Finally, note that these continuum contributions enter as an independent 
input in our analysis. They are not fitted to the form factor data.

\section{Spectral Functions}
\label{sec:spec}

\subsection{Structure}
\label{sec:specf}

As discussed above, the spectral function can at present only be obtained
from unitarity arguments and experimental data for 
the lightest two-particle intermediate states ($2\pi$ and $K\bar{K}$)
\cite{HP,Hammer:1998rz,Hammer:1999uf}.
The $\rho\pi$ continuum contribution has been calculated in the Bonn-J\"ulich
$NN$ model \cite{Meissner:1997qt}.

The remaining contributions to the spectral function
can be parameterized by vector meson poles. On one hand, 
the lower mass poles can be identified with physical vector 
mesons such as the $\omega$ and the $\phi$. In the the case of the 
$3\pi$ continuum, e.g., it has been shown in ChPT that the nonresonant
contribution is very small and the spectral function is dominated 
by the $\omega$ \cite{BKMspec}.
The higher mass poles on the other hand, are simply an effective way
to parameterize higher mass strength in the spectral function.
Different parameterizations are possible and an explicit example will
be discussed below in relation to the pQCD behavior.

%
%

In all our fits the spectral function includes the 
$2\pi$, $K\bar{K}$, and  $\rho\pi$ continua from unitarity
and the  $\omega$ pole. Note that we also include a pole at the $\phi$ 
mass to account for explicit $\phi$ strength not included in 
the $K\bar{K}$ and $\rho\pi$ continua.
In addition to that there are a number
of effective poles at higher momentum transfers
in the isoscalar ($s_1,s_2,...$) and isovector 
channels ($v_1,v_2,...$).
The spectral function has the general structure
\bea
{\rm Im }\,F_i^{s} (t) &=& {\rm Im }\,F_i^{(s,K\bar{K})} (t)
+ {\rm Im }\,F_i^{(s,\rho\pi)} (t) 
+ \sum_{V=\omega,\phi,s_1,...} \pi a_i^{V}
\delta (M^2_{V}-t) \,, \quad i = 1,2 \, ,
\label{emff:s} \\
{\rm Im }\,F_i^{v} (t) &=& {\rm Im }\,F_i^{(v,2\pi)} (t)
+ \sum_{V= v_1,...} \pi a_i^{V}
\delta (M^2_{V}-t) \,, \quad i = 1,2 \, .
\label{emff:v}
\eea

The masses of the effective poles are fitted to the form factor data. 
We generally do not include widths for the effective poles. However,
in some of the fits we allow a large width for the highest mass effective
pole in order to mimick the imaginary part of the form factors in the 
time-like region. We have performed various fits with 
different numbers of effective poles and including/excluding some of the 
continuum contributions. In Sec.~\ref{sec:fits}, we will discuss
the results of these efforts.

\subsection{Constraints}
\label{sec:con}

The number of parameters in the fit function is reduced by enforcing
various constraints.
The first set of constraints concerns the low-$t$ behavior of the
form factors: We enforce the correct normalization of the form factors
as given in Eq.~(\ref{norm}). The nucleon radii, however, are not 
included as a constraint. In some earlier fits, we had also 
constrained the neutron charge radius to the value from 
low-energy neutron-atom scattering experiments \cite{Kop95,Kop97}. 
In the fits discussed below, this constraint is dropped since the fit 
value is compatible with the empirical range from Refs.~\cite{Kop95,Kop97}.

Perturbative QCD (pQCD) constrains the behavior of the nucleon
electromagnetic form factors for large momentum transfer.
Brodsky and Lepage \cite{BrL80} find for $Q^2 \to \infty$,
\begin{equation}
F_i (t) \to \frac{1}{Q^{2(i+1)}} \, 
\left[ \ln\left(\frac{Q^2}{Q_0^2}\right)
\right]^{-\gamma} \, , \quad i = 1,2 \, ,
\label{emff:fasy1}
\end{equation}
where $Q_0 \simeq \Lambda_{\rm QCD}$.
The anomalous dimension $\gamma\approx 2$ depends weakly on the number of
flavors \cite{BrL80}.
The asymptotic behavior of the form factors has recently also been studied 
in connection to the unexpected behavior of the ratio 
$Q^2 F_2 (Q^2)/F_1(Q^2)$ for the proton
measured at Jefferson Lab 
and different expressions for the logarithmic corrections were
found \cite{Belitsky:2002kj,Ralston:2003mt}.
(For a further discussion of the asymptotic behavior of the nucleon
form factors for large space-like and time-like momenta, see
Ref.~\cite{Tomasi-Gustafsson:2005sk}.)
In the current analysis we implement only the leading power
behavior of the form factors and details of the logarithmic
correction are not relevant.
Note that the logarithmic term in Eq.~(\ref{emff:fasy1}) was included in
some of our earlier analyses \cite{MMD96,HMD96,HM04} but had
little impact on the fit. The particular way this constraint was
implemented, however, lead to an unphysical logarithmic singularity 
of the form factors in the time-like region which we want to avoid in the
current analysis.

The power behavior of the form factors at large $t$ can be easily 
understood from perturbative gluon exchange. In order to distribute the 
momentum transfer from the virtual photon
to all three quarks in the nucleon, at least two massless
gluons have to be exchanged. Since each of the gluons has a propagator 
$\sim 1/t$, the form factor has to fall off as $1/t^2$. In the case
of $F_2$, there is additional suppression by $1/t$ since a quark spin 
has to be flipped.
The power behavior of the form factors
leads to superconvergence relations of the form
\begin{equation}
\label{eq:scr}
\int_{t_0}^\infty {\rm Im}\, F_i (t) \;t^n dt =0\, , \quad i = 1,2 \, ,
\end{equation}
with $n=0$ for $F_{1}$ and $n=0,1$ for $F_{2}$.

The pQCD power behavior can be enforced in various ways. To obtain
some information about the induced theoretical uncertainty, we
discuss two different methods in more detail:
\begin{enumerate}
\item Superconvergence (SC) approach:

The asymptotic behavior of Eq.~(\ref{emff:fasy1}) is obtained by
choosing the residues of the vector meson pole terms such that the
leading terms in the $1/t$-expansion cancel. This leads to a 
spectral function consistent 
with the superconvergence relations Eq.~(\ref{eq:scr})
and the asymptotic behavior Eq.~(\ref{emff:fasy1}). 
This method is similar to what was used in earlier works 
\cite{MMD96,HMD96,HM04}. Here we add
a very broad resonance of the structure
\beq
\label{eq:broad}
F^{(I,broad)}_i(t) = \frac{a^I_i (M_{I}^2-t)}{(M_{I}^2-t)^2
+\Gamma_{I}^2}\,,\quad i = 1,2 \, , \quad I = s,v \, ,
\eeq
in both the isovector and isoscalar form factors.  
The resonance parameterises continuum contributions in addition to the 
$2\pi$, $K\bar{K}$, and $\rho\pi$ continua and generates an imaginary
part of the form factors in the time-like region for $t\geq 4 M^2$.
The residue $a^I_i$ as well as the mass and width parameters $M_{I}$ and
$\Gamma_{I}$ are fit to the data. The width parameter
$\Gamma_{I}$ is of the same order of magnitude as the mass $M_{I}$ and
comes out typically of the order of a few GeV in our fits.

\item Explicit pQCD continuum approach:

In addition to satisfying the superconvergence relations, Eq.~(\ref{eq:scr}),
a term of the form
\beq
\label{eq:expQCD}
F_i^{(I,pQCD)} = \frac{a_{i}^{I}}{1-c_{i}^2 t+b_{i}^2 (-t)^{i+1}}\,,
\quad i = 1,2 \, , \quad I = s,v \, ,
\eeq
which explicitly enforces the pQCD behavior, Eq.~(\ref{emff:fasy1}),
is added to the fit function. 
Such a term behaves like an effective resonance pole for small values of $t$, 
and restores pQCD behaviour explicitly at high values
of $t$. The superconvergence relations cancel the leading order terms in the
$1/t$ expansion. This explicit pQCD term is consistent with a 
nonvanishing imaginary part of the form factors in the time-like region.
Note that the parameters $b_{i}$ and $c_{i}$ are the same
in the isoscalar and isovector channels while the residue $a_{i}^{I}$
depends on the channel.
This method allows for a smoother interpolation between the low-$t$ and
large-$t$ regions compared to the SC approach. Because of this 
feature, one might expect obtaining fits with fewer parameters.

\end{enumerate}

The number of effective poles in
Eqs.~(\ref{emff:s}, \ref{emff:v}) is determined
by the stability criterion discussed in detail in \cite{Sab80}.
In short, we take the minimum number of poles necessary to fit the data.
The number of free parameters is strongly reduced by the 
various constraints (unitarity, normalizations, superconvergence
relations). More details will be given together with
the fits in the next section.

\section{Fit Results}
\label{sec:fits}

The fits have been performed using the Fletcher-Reeves and the Polak-Ribiere
conjugate gradient algorithms implemented in the GSL library~\cite{GSL}. To 
ensure initial convergence stability, Monte-Carlo sampling was performed over 
the whole physically acceptable parametric volume to obtain a number of
parameter sets with acceptable starting $\chi^{2}$ values.

The constraints dictated by the normalization and the pQCD conditions have
been represented in terms of a set of linear equations for the resonance
residua. The equations are solved each iteration using the LU
decomposition.

Soft constraints on composite variables which depend on a set of fit
parameters allow to impose an exponential well for the set of parameters
as a whole, limiting deviations of the composite variable from its desired
central value. These constraints are implemented as additive $\chi^{2}$
terms of the general form
\beq
\tilde{\chi}^{2} = p [x-\langle x \rangle]^{2} 
\exp(p [x-\langle x\rangle]^{2})
\eeq
where $\langle x \rangle$ is the desired central value, and $p$ is the
constraint strength parameter which allows to stabilize fit convergence over
the whole range of iterations adaptively, regulating the steepness of the
exponential well. This method was used in some earlier fits to constrain
the neutron radius \cite{Hammer:2006mw}.

The error bands are obtained by allowing the total $\chi^2$/dof
of the fit to be in the interval $[\chi^2_{min}, \chi^2_{min}+\delta\chi^{2}]$
where $\chi^2_{min}$ is the $\chi^2$ value of the best fit, and
$\delta\chi^{2}$ is obtained from the $1\sigma$ 
confidence interval $p$-value equations, $\delta\chi^{2}\simeq 1.04$.

The data basis used in the fits is taken from Ref.~\cite{FW} and in addition 
also includes the new data that have appeared since 2003 and the time-like 
data~\cite{Plaster:2005cx,Glazier:2004ny,Warren:2003ma,Bermuth:2003qh,Ziskin:Thesis,unknown:2006jp,Ambrogiani:1999bh,Ablikim:2005,Bardin:1994am,Antonelli:1994kq,Bassompierre:1977ks,Delcourt:1979ed,Bisello:1990rf,Pedlar:2005sj,Aubert:2005cb,Antonelli:1993vz,Antonelli:1998fv}.
The  CLAS collaboration at Jefferson Lab has performed measurements of the 
neutron magnetic form factor for momentum transfers $0.6 \leq Q^2 \leq
5\,$GeV$^2$ \cite{CLASdata}. These data are still preliminary and are 
therefore generally not included in our fits. In subsection \ref{sec:pCLAS},
however, we discuss a fit where these preliminary data are included.
The results for $G_{M}^{n}$, $G_{E}^{p}$, $G_{M}^{p}$ are normalized to the
phenomenological dipole fit:
\beq
\label{dipole}
G_D( Q^2)=\left(1+\frac{Q^2}{m_D^2}\right)^{-2} \,,
\eeq
where $m_D^2=0.71$ GeV$^2$.

Additionally, certain features of the form factor behaviour in specific
$Q^{2}$ ranges can be enhanced during the fitting procedure by artificially
decreasing the errors on the experimental data in that region as seen by the
fit. This allows to, for example, explore the conditions necessary to
produce a pronounced bump-dip structure in $G_{E}^{n}$, discussed in
Sec.~\ref{sec:picloud}.

In the time-like region, the neutron data do not participate in the fit as
they are obtained from a single experiment. They are therefore a genuine
prediction.

\subsection{Superconvergence (SC) Approach}
\label{sec:SCA}

\begin{figure*}[ht] 
\centerline{\includegraphics*[width=15cm,angle=0]{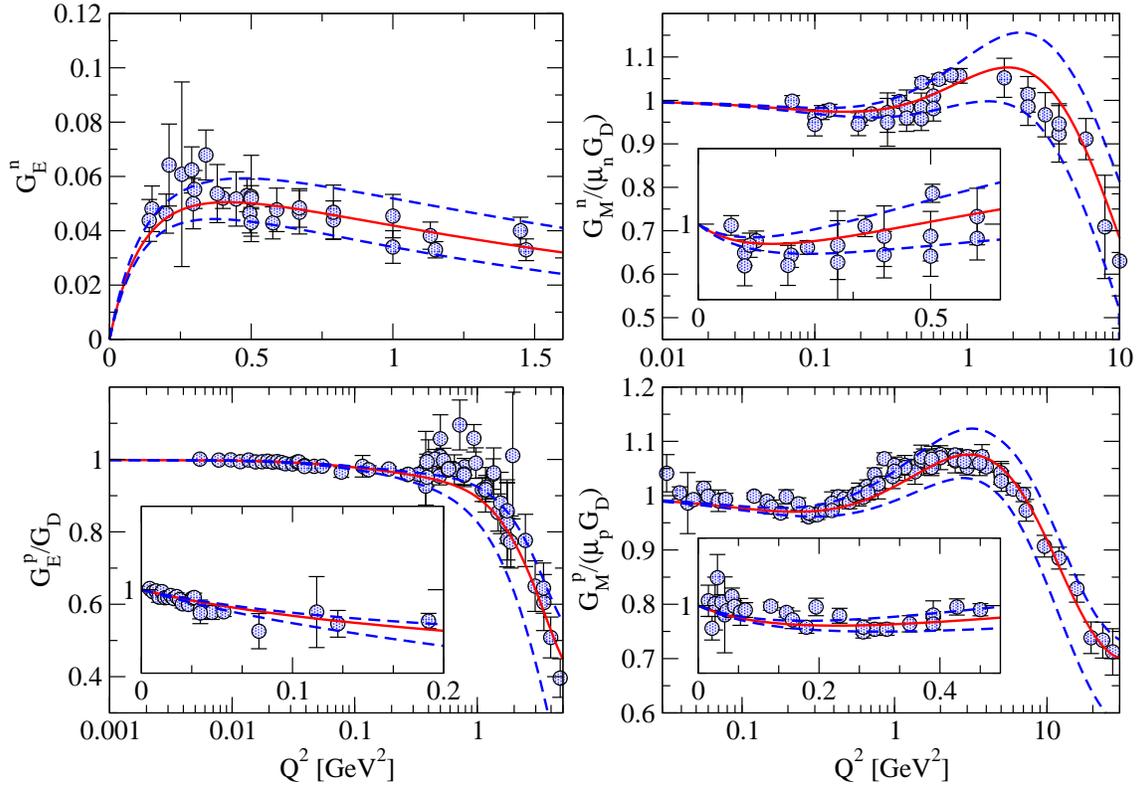}}
\caption{The nucleon electromagnetic form factors for space-like
momentum transfer in the SC approach.
The solid line gives our best fit while the dashed lines indicate the error
band obtained by the $1\sigma$ deviation as discussed above.
}
\label{fit10}
\end{figure*} 
In Fig.~\ref{fit10}, we show the results in the 
SC approach for all four form factors 
in the space-like region compared to the world data.
In general, we get a good description of all data within
our error bands. For $G_E^p$ there is some inconsistency in the data
points around $Q^2\approx 1$ GeV$^2$. Our fit favors the lower data
points in this region. As can be seen from the inset, the 
data at low momentum transfers are well described.
In addition to the $\omega$ and the residual $\phi$, this fit has 2 
more isoscalar poles ($M_{s_1}\approx 1.1$ GeV, $M_{s_2}\approx 2.0$ GeV)
and 5 isovector poles with masses ranging from 1 to 3 GeV.
The heaviest poles in both channels are broad resonances 
(cf.~Eq.~(\ref{eq:broad})) with width parameters
ranging from 5 GeV (isoscalar)
to 19 GeV (isovector). All other poles have
zero widths. The fit has 17 free parameters and a total $\chi^2$/dof of 1.8.  
The fit parameters are listed in detail in Table \ref{tab:parsc}
in the Appendix.

Note also that we do not obtain a pronounced bump-dip structure in
$G_E^n$ as observed in  Ref.~\cite{FW}. However, all data for 
$G_E^n$ are described within our error band and the experimental
error. We will come back to this bump-dip structure
in Sec.~\ref{sec:picloud} and discuss the modifications in the spectral
function required to produce this structure.

\begin{figure*}[ht]
\centerline{\includegraphics*[width=15cm,angle=0]{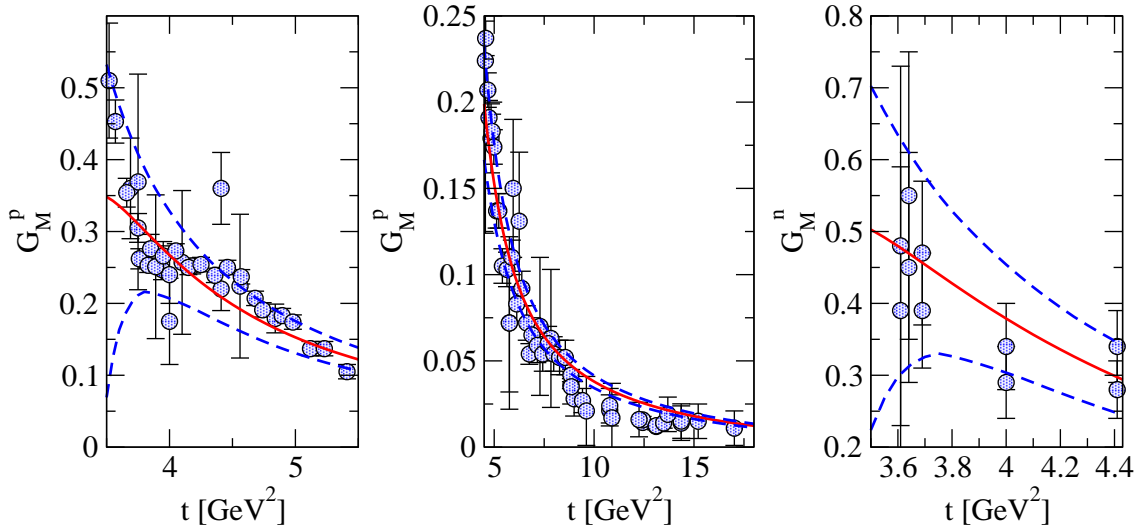}}
\caption{The nucleon electromagnetic form factors for time-like
momentum transfer in the SC approach
(right panel: $G^M_n$, left panels: $G^M_p$). 
The proton data participate in the fit while
the neutron data are a genuine prediction. 
The solid line gives our best fit while the dashed lines indicate the error
band obtained by the $1\sigma$ deviation as discussed above.
}
\label{tfit10}
\end{figure*}

In Fig.~\ref{tfit10}, we show our fit results for the SC approach
in the time-like region.
As in the space-like region we get a good description of the world data
within our error bands. Our best fit, however, cannot reproduce the
strong rise of the data for $G_M^p$ near threshold. The data for 
$G_M^n$ are also well described. Note that the neutron data do not
participate in the fit and the corresponding curves are therefore
genuine prediction of the dispersion analysis based on data in the
other channels.

\subsection{Explicit pQCD Continuum Approach}
\label{sec:expQCD}

\begin{figure*}[ht] 
\centerline{\includegraphics*[width=15cm,angle=0]{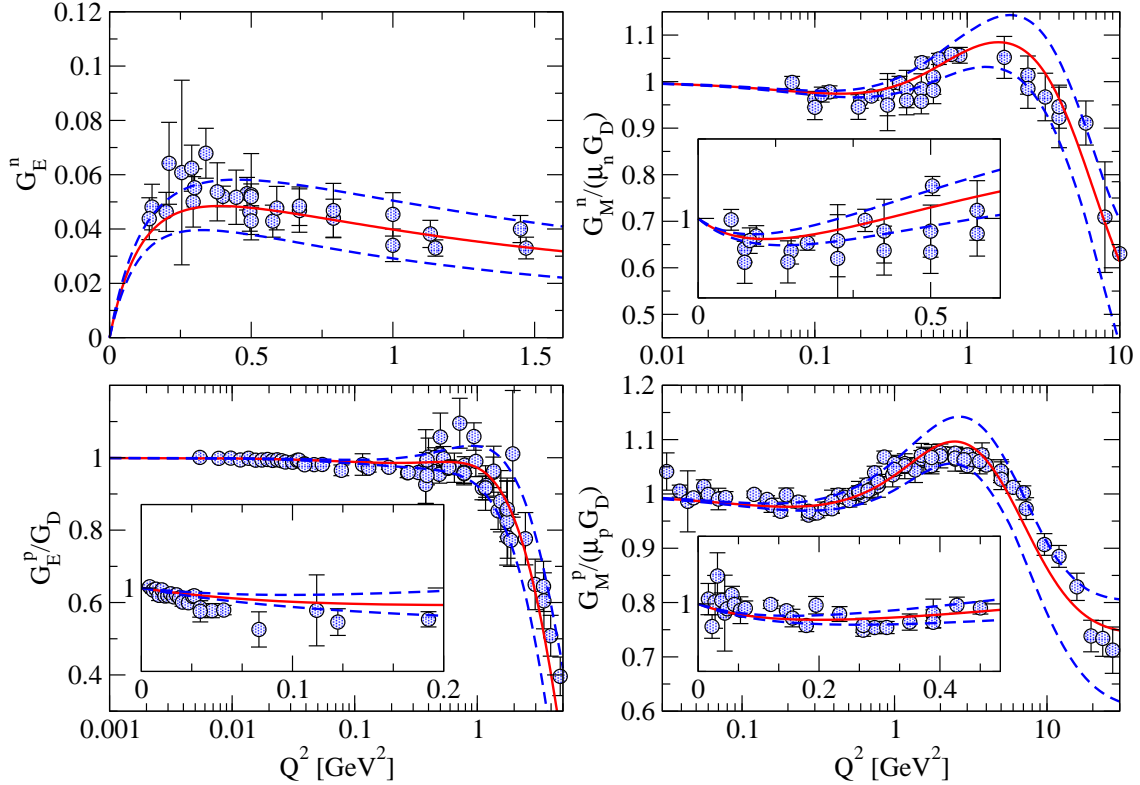}}
\caption{The nucleon electromagnetic form factors for space-like
momentum transfer with the explicit pQCD continuum.
The solid line gives our best fit while the dashed lines indicate the error
band obtained by the $1\sigma$ deviation as discussed above.
}
\label{fit14}
\end{figure*}

In Fig.~\ref{fit14}, we show our results in the 
explicit pQCD continuum approach for space-like 
momentum transfers compared to the world data.
Again, we get a good description of all data within
our error bands. In contrast to the superconvergence approach, our
fit now favors somewhat larger values of  $G_E^p$ in   
the region around $Q^2\approx 1$ GeV$^2$.
For $G_E^n$ the situation is the same as before: we describe all 
data within our error band and the experimental
errors but see no pronounced bump-dip structure in the fits.
In addition to the $\omega$ and the residual $\phi$, this fit has 
one more isoscalar pole ($M_{s_1}\approx 1.8$ GeV)
and 3 isovector poles ($M_{v_1}\approx 1.0$ GeV, $M_{v_2}\approx 1.6$ GeV,
and  $M_{v_3}\approx 1.8$ GeV). Moreover, it contains an explicit pQCD
continuum term, Eq.~(\ref{eq:expQCD}), as discussed above.
This fit has 14 free parameters and a total $\chi^2$/dof of 2.0.  
In general, we cannot get a satisfactory description of all data
using fewer parameters. The parameters of this fit are 
given in detail in Tables \ref{tab:parexpQCD1} and \ref{tab:parexpQCD2}
in the Appendix.

In Fig.~\ref{tfit14}, we show the results in the time-like region. 
\begin{figure*}[ht] 
\centerline{\includegraphics*[width=15cm,angle=0]{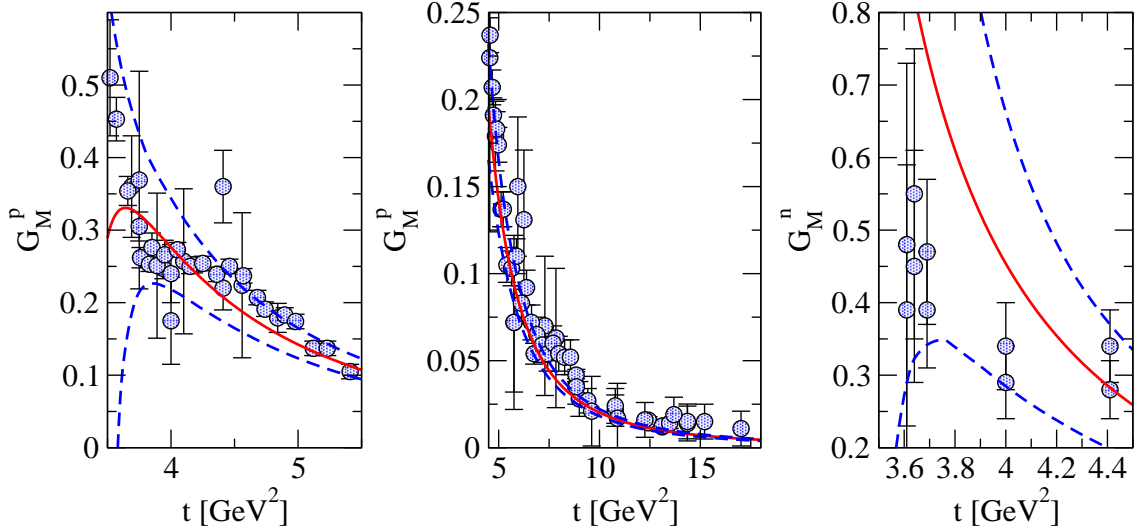}}
\caption{The nucleon electromagnetic form factors for time-like
momentum transfer with the explicit pQCD continuum
(right panel: $G^M_n$, left panels: $G^M_p$). 
The proton data participate in the fit while
the neutron data are a genuine prediction. 
The solid line gives our best fit while the dashed lines indicate the error
band obtained by the $1\sigma$ deviation as discussed above.
}
\label{tfit14}
\end{figure*} 
As before we describe the time-like data within our error band, but
the errors increase strongly close to threshold. Our best
fit turns over very close to threshold and cannot describe the two
lowest data points. The error band is even larger for the neutron data
which do not participate in the fit. But within the $1\sigma$ band the
neutron data are well described by this fit. Due to the strong increase
of the $1\sigma$ band, however, we can not
make precise predictions for the time-like form factors close to threshold.

\subsection{Nucleon Radii and Coupling Constants}

\label{sec:nucrad}

In Table \ref{tab:nucrad}, we give the nucleon radii extracted from
our fits in the SC and explicit pQCD approaches.
\begin{table}[ht]
\begin{tabular}{|c|c|c|c|c|}
\hline 
& SC approach & explicit pQCD app. & Ref.~\cite{HM04} & recent determ.\\
\hline\hline
$r_E^p$ [fm] & 0.844 ($0.840\,\ldots\,0.852$) & 0.830 ($0.822\,\ldots\,0.835$)  &0.848 & 0.886(15) 
\cite{Rosenfelder:1999cd,Sick:2003gm,Melnikov:1999xp}\\
$r_M^p$ [fm] & 0.854 ($0.849\,\ldots\,0.859$)& 0.850 ($0.843\,\ldots\,0.852$)  &0.857 & 0.855(35) 
\cite{Sick:2003gm,Sick:private} \\
$(r_E^n)^2$ [fm$^2$] & $-0.117$ ($-0.11\,\ldots\,-0.128$) & $-0.119$
($-0.108\,\ldots\,-0.13$)  & $-0.12$ & $-0.115(4)$ \cite{Kop97} \\
$r_M^n$ [fm] & 0.862 ($0.854\,\ldots\,0.871$) & 0.863 ($0.859\,\ldots\,0.871$)  &0.879 & 0.873(11) \cite{Kubon:2001rj}\\
\hline
\end{tabular}
\caption{\label{tab:nucrad}
Nucleon radii extracted from our fits in the SC (2nd column) and 
explicit pQCD approaches (3rd column). 
The first number gives the value for our best fit, while the 
numbers in parentheses indicate the  range from the $1\sigma$ band.
For comparison, we give the results
of Ref.~\cite{HM04} (4th column) and other recent determinations from
low-momentum transfer data
\cite{Rosenfelder:1999cd,Sick:2003gm,Melnikov:1999xp,Sick:private,Kop97,Kubon:2001rj}
(5th column).}
\end{table}
The first number gives the value for our best fit, while the 
numbers in parentheses indicate the  range from the $1\sigma$ band.
Our values are compared to the results
of Ref.~\cite{HM04} and other recent determinations from
low-momentum transfer data
\cite{Rosenfelder:1999cd,Sick:2003gm,Melnikov:1999xp,Sick:private,Kop97,Kubon:2001rj}.

The nucleon radii are generally in good agreement with
other recent determinations using only low-momentum-transfer data 
given in the table. In particular, the squared neutron charge radius 
is in good agreement with the experimental value.
In previous analyses \cite{MMD96,HM04}, this radius was constrained
to the experimental value and not a prediction.
Our result for the proton charge radius, however,
is somewhat small. This was already the case in the earlier dispersion
analyses of Refs.~\cite{MMD96,HM04}. 
We speculate that the reason for this discrepancy
lies in inconsistencies in the data sets. In this type of global analysis
all four form factors are analyzed simultaneously and both data
at small and large momentum transfers enter. This can be an advantage
or a disadvantage depending on the question at hand.

The discrepancy is not likely to be explained by $2\gamma$ physics.
In Ref.~\cite{Sick05}, it was shown that $2\gamma$ exchange
(including intermediate nucleons only) has a very tiny effect on the 
extraction of the proton radius from $ep$ scattering data.
We have also performed various fits where the proton charge radius was 
constrained to values between 0.88 and 0.90~fm. The quality of these
fits is not quite as good as for our best fits
but still acceptable ($\chi^2$/dof~$\approx 2.6$).
The fits exhibit a pronounced bump-dip structure in $G_E^p$. However, 
they are on the low side of the data in the interval $Q^2=0.3...0.8$ 
GeV$^2$. These conclusions remain valid if we use a 
$2\gamma$ corrected data basis for the proton form factor
(including intermediate nucleons in the $2\gamma$ exchange only)
\cite{ArringtonP}.

We have also extracted the $\omega NN$ couplings from our fit.
The $\omega NN$ coupling constant is related to the residua via
\beq
g^i_{\omega NN} =\frac{f_\omega}{M_\omega^2}\, a_i^\omega\,,\quad
i=1,2\,,
\eeq
where $f_\omega=17$ is the electromagnetic coupling of the $\omega$.
The $\omega$ resonance couplings, e.g., play an important role in
addressing the issue of isospin violation in the nucleon form factors
\cite{Kubis:2006cy}. While the vector residua of the 
$\omega$ are fixed relatively well by the fits, 
$a_{\omega}^{1} = 0.60\,\ldots\,0.83$~GeV$^2$,
even the sign of the tensor residua can not be determined, $a_{\omega}^{2} =
-0.13\,\ldots\,0.37$~GeV$^2$.
This leads to the following range for the $\omega NN$ coupling constants:
\beq
g^1_{\omega NN} = 16.7...23.1\,,\quad \mbox{ and }
\quad g^2_{\omega NN} = -3.6...10.3\,.
\eeq
We have not extracted a coupling constant for the $\phi$ to the nucleon
since the $\phi$ strength appears also in the $K\bar{K}$ and $\rho\pi$
continua and the interpretation of the residual $\phi$ pole strength is
ambiguous. 

\subsection{Inclusion of the preliminary CLAS data}
\label{sec:pCLAS}

\begin{figure*}[ht] 
\centerline{\includegraphics*[width=15cm,angle=0]{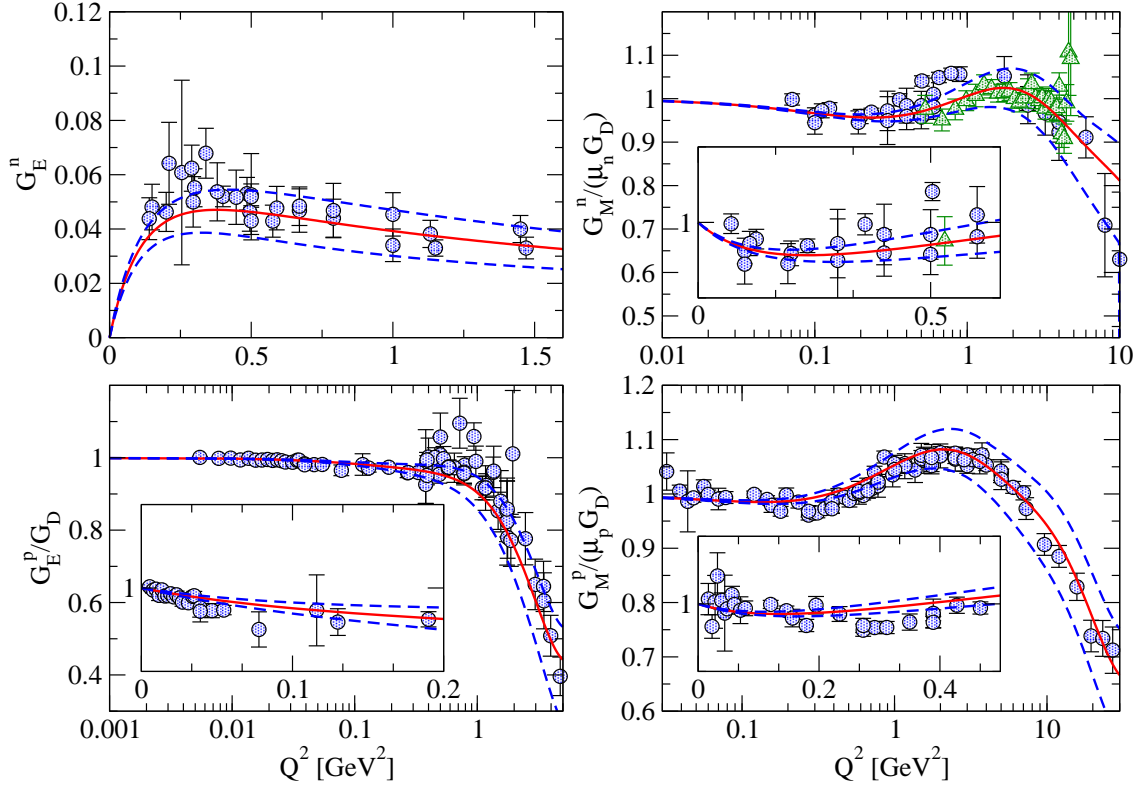}}
\caption{The nucleon electromagnetic form factors for space-like
momentum transfer with the explicit pQCD continuum. The blue circles 
indicate the published world data while the green triangles give
the preliminary CLAS data for $G_M^n$ \cite{CLASdata}.
The solid line gives our best fit while the dashed lines indicate the error
band obtained by the $1\sigma$ deviation as discussed above.
}
\label{fit13}
\end{figure*}

The CLAS collaboration at Jefferson Lab
has recently taken new data for $G_M^n$
in the range $0.6 \leq Q^2 \leq 5\,$GeV$^2$ \cite{CLASdata}. 
These data are still preliminary and were not included in the fits discussed
above. In the this subsection, we present a fit  in the 
explicit pQCD continuum approach where the preliminary CLAS data are
included. In Fig.~\ref{fit13}, we show our results for space-like 
momentum transfers compared to the published 
world data (blue circles)
and the preliminary CLAS data (green triangles) \cite{CLASdata}.
Again, we get a good description of most data within
our error bands. In the region $0.5$~GeV$^2$ $\simlt Q^2 \simlt 1.0$~GeV$^2$ 
the preliminary CLAS data differ significantly 
from most of the published world data. 
The reason for this discrepancy is not yet understood. Our fit
prefers the CLAS data in this region.
Some of the world data are even out of our  $1\sigma$ band.
The description of the timelike data and the nucleon radii
in this fit is of similar quality as for  the fits 
described in subsections \ref{sec:SCA} and \ref{sec:expQCD}.

In addition to the $\omega$, this fit has 
two more isoscalar poles ($M_{s_1}\approx 1.1$~GeV
and $M_{s_2}\approx 1.4$~GeV)
and 3 isovector poles ($M_{v_1}\approx 1.0$~GeV, $M_{v_2}\approx 1.6$~GeV,
and  $M_{v_3}\approx 1.8$~GeV). Moreover, it contains an explicit pQCD
continuum term, Eq.~(\ref{eq:expQCD}), as discussed above.
This fit has 15 free parameters and a total $\chi^2$/dof of 2.2.  
The slight increase in the total $\chi^2$ compared to the similar fit in 
subsection \ref{sec:expQCD} is due to the inconsistency between the
preliminary CLAS data and the older data in
the region $0.5$~GeV$^2$ $\simlt Q^2 \simlt 1.0$~GeV$^2$.
The parameters of this fit are 
given in detail in Tables \ref{tab:parexpCLAS1} and \ref{tab:parexpCLAS2}
in the Appendix.

\subsection{Pion Cloud of the Nucleon}
\label{sec:picloud}

Friedrich and Walcher (FW) recently analysed the em nucleon 
form factors and performed
various phenomenological fits \cite{FW}. Their fits showed a pronounced
bump-dip structure in $G_E^n$ which they interpreted using an ansatz for the 
pion cloud based on the idea that the proton can be thought of as a
virtual neutron-positively charged pion pair. 
They found a very long-range contribution to the charge distribution 
in the Breit frame extending out to about 2~fm which they attributed to 
the pion cloud. 
While naively the pion Compton wave length is of this size, 
these findings are indeed surprising if compared with the ``pion cloud'' 
contribution due to the $2\pi$ continuum
contribution to the isovector spectral 
functions discussed in Sec.~\ref{sec:2pico}.

As was shown by Hammer, Drechsel, and Mei\ss ner \cite{Hammer:2003qv},
the $2\pi$ continuum contributions to  the long-range part of the nucleon 
structure are much more confined in coordinate space
and agree well with calculations in ChPT \cite{Norbert} and
earlier (but less systematic) calculations based on chiral 
soliton models, see e.g. \cite{UGM}. 
In the dispersion-theoretical framework, the longest-range part of the 
pion cloud contribution to the nucleon form factors is given
by the $2\pi$ continuum -- the lowest-mass intermediate 
state including only pions.

As a consequence, it remains to be shown
how the proposed long-range pion cloud can be reconciled with
what is known from dispersion relations and ChPT.
In order to clarify this issue, we have performed various fits in 
order to understand what structures in the spectral function are required to
reproduce the bump in $G_E^n$.
We find that the structure can only be reproduced if additional low-mass
strength in the spectral function below $t \simlt 1$ GeV$^2$ is allowed
beyond the $2\pi$, $K\bar{K}$, and $\rho\pi$ continua and the 
$\omega$ pole. (See also Ref.~\cite{Hammer:2006mw} for some preliminary 
results on this question.) Since the spectral function is
well understood in this region in terms of meson continua and vector meson
dominance, such strength was explicitly excluded
in the fits of Sec.~\ref{sec:fits}.

\begin{figure*}[ht] 
\centerline{\includegraphics*[width=15cm,angle=0]{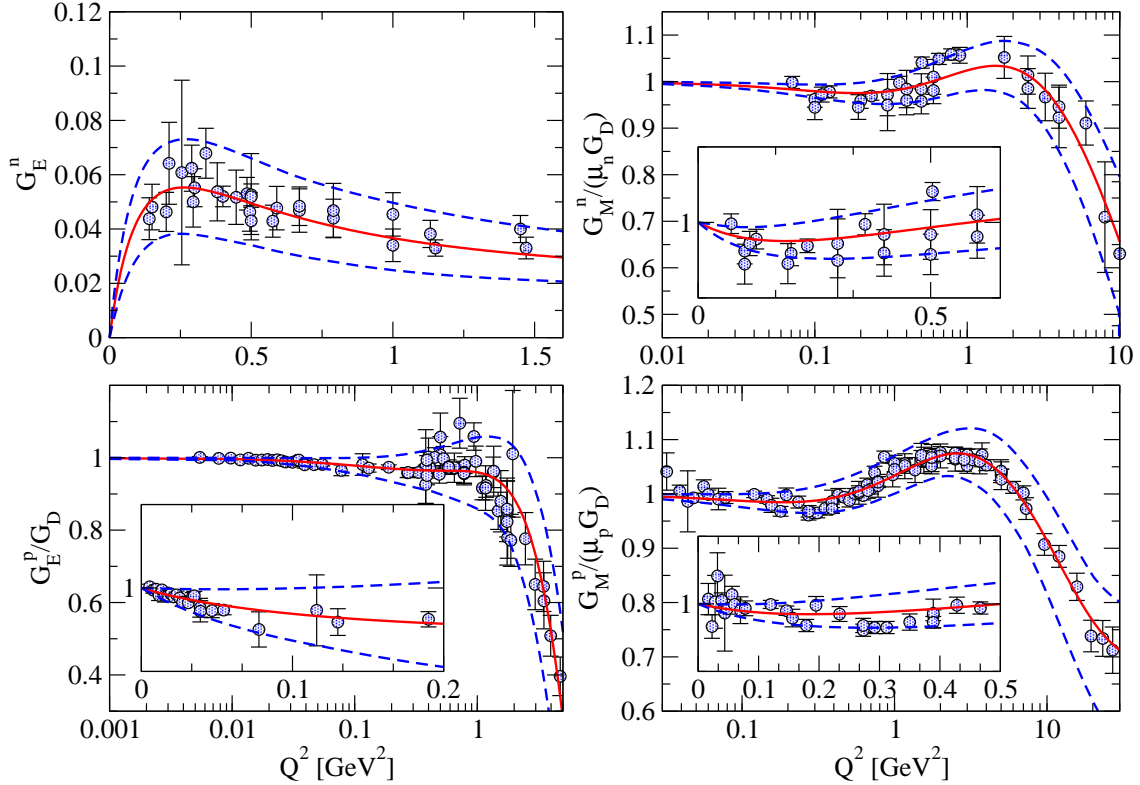}}
\caption{The nucleon electromagnetic form factors for space-like
momentum transfer with bump-dip structure for $G_E^n$.
The solid line gives our best fit while the dashed lines indicate the error
band obtained by the $1\sigma$ deviation as discussed above.
}
\label{fit16}
\end{figure*} 
In Fig.~\ref{fit16}, we show the results in the explicit pQCD
approach with additional low-mass strength allowed. 
In this fit, all constraints were removed, the neutron charge form factor
behaviour in the region of the bump-dip
structure was enhanced by artificially lowering
the error bars on the experimental data as seen by the fit. 
The results for all four form factors 
in the space-like region are compared to the world data.

In general, we get a good description of all data within
our error bands. In particular, we obtain the desired bump-dip structure
in $G_E^n$
at the cost of low-mass poles, which appear close to the $\omega$ mass in the
isoscalar channel, and close to the 3-pion threshold in the isovector
channel. The latter pole is weakly coupled. The behavior of the 
proton charge form factor around $Q^2\sim 1$ GeV$^2$ 
is not resolved. The fit 
has 22 parameters and two effective poles (one isoscalar and one isovector)
come out below 1 GeV. In addition to the $\omega$, this
fit has 3 isoscalar poles
and 4 isovector poles. One of the isovector poles is a broad
resonance with a width parameter of order 13 GeV.
The time-like data are not included in the
fit. For the space-like data alone the total $\chi^2$/dof is 0.9. 
The parameters of this fit are 
listed in detail in Tables \ref{tab:parFW1} and \ref{tab:parFW2}
in the Appendix.

While this fit gives a good description of the space-like data,
it is in contradiction to much of what is known about the structure of
the spectral function below 1~GeV. 
Finally, we note again that the error bars of the data on $G_E^n$
in the region of the bump-dip structure had to be lowered artificially
in order to obtain the desired structure. Taking it seriously requires
to overcome the constraints imposed by unitarity and analyticity in 
the  spectral function below 1~GeV. 

\section{Summary \& Outlook}
\label{sec:summ}
Dispersion theory simultaneously describes all four nucleon 
form factors over the whole range of momentum transfers in both the 
space-like and time-like regions. It allows for the inclusion of
constraints from other physical processes, unitarity, and ChPT
and therefore is an ideal tool to analyze the form factor data.

We have presented the results of our new form factor analysis.
The spectral function has been improved compared to earlier
analysis in various respects.
(i) It contains the updated $2\pi$ continuum 
\cite{Belushkin:2005ds}, as well the $K\bar{K}$ 
\cite{Hammer:1998rz,Hammer:1999uf} and $\rho\pi$ 
continua \cite{Meissner:1997qt} as independent input.
(ii) The pQCD behavior of the form factors at large momentum transfer
has been included in two different ways: using superconvergence
relations and a broad resonance to mimick the QCD continuum (SC) and
by including an explicit pQCD continuum term.
Moreover, we have generated $1\sigma$ error bands for all our fits by 
performing a Monte Carlo sampling of all fits with a total
$\chi^2$/dof in the interval $[\chi_{min}^2, \chi_{min}^2+1.04]$.

Our fits give a consistent description of the world data in the 
space-like  and time-like regions. 
We find that some residual $\phi$ pole strength is still
required in addition to the $K\bar{K}$ and $\rho\pi$ 
continua.
The FENICE data for the neutron
time-like form factors do not participate in the fit but are 
reproduced within the errors bands. This is an improvement
over our earlier studies \cite{HMD96,slacproc}.
The nucleon radii are generally in
good agreement with other determinations with the exception
of the proton charge radius which comes out smaller for 
our best fits. As discussed in subsection \ref{sec:nucrad}, enforcing 
a larger radius $r_E^p \simeq 0.88$ fm leeds to acceptable fits
with a slightly higher $\chi^2$. 
While the $\omega NN$ vector coupling constant is determined 
relatively well by the fits, even the sign of the tensor coupling constant
can not be determined. We have also performed fits including the
preliminary data for $G_M^n$ from the CLAS collaboration at Jefferson Lab 
\cite{CLASdata}.  For lower $Q^2$ these data are
in conflict with earlier data, still our fits seem to
prefer the trend set by the CLAS data. 
The bump-dip structure in $G_E^n$ advocated
by Friedrich and Walcher \cite{FW} can only be obtained by artificially
lowering the error bars of the data and allowing additional strength in the
spectral function below 1 GeV.

There are two important extensions of this work. First, one could fit
directly to the Coulomb-corrected cross section data, which would
eliminate possible inconsistencies in the form factor data basis
\cite{Sick:private}. This could be further refined by consistently
removing the full $2\gamma$ exchange from the cross section data.
The latter, however, remains a formidable challenge because of the 
contribution of the nucleon excited states.

\section*{Acknowledgments}

We thank  Bastian Kubis and Ingo Sick for useful discussions.
We also thank John Arrington, Will Brooks and Richard Milner for
informative communications and the CLAS collaboration for
allowing us to show their data prior to publication.
The work was supported in part by the EU I3HP 
\lq\lq Study of Strongly Interacting Matter'' under contract number 
RII3-CT-2004-506078 and the DFG through funds provided
to the SFB/\-TR 16 \lq\lq Subnuclear Structure of Matter''.

\newpage
\appendix

\section{Fit Parameters}

In this Appendix, we give the values of the fit parameters for the
various fits.
Table \ref{tab:parsc} contains the parameter values for the fit in the 
SC approach.
\begin{table}[htb]
\begin{tabular}{|c|c|c|c|c|}
\hline
Resonance & Mass [GeV] & $a_1$ [GeV$^{2}$] & $a_2$ [GeV$^{2}$] & $\Gamma$ 
[GeV]\\
\hline
\hline
$\omega$ & 0.782 & 0.755960 & 0.370592 & $-$ \\ 
\hline
$\phi$ & 1.019 & $-0.776537$ & $-2.913229$ & $-$\\
\hline
$s_1$ & 1.124860 & 0.902379 & 2.484859 & $-$\\
\hline
$s_2$ & 2.019536 & 0.022798 & $-0.130622$ & 5.158635 \\
\hline
\hline
$v_1$ & 1.062128 & $-0.127290$ & $-2.162533$ & $-$\\
\hline
$v_2$ & 1.300946 & $-1.243412$ & 3.704233 & $-$\\
\hline
$v_3$ & 1.493630 & 4.191380 & $-7.091021$ & $-$\\
\hline
$v_4$ & 1.668522 & $-3.176013$ & 3.723858 & $-$\\
\hline
$v_5$ & 2.915451 & 0.048987 & 0.075965 & 19.088297 \\
\hline
\end{tabular}
\caption{\label{tab:parsc}
Fit parameters in the SC approach.
This fit has 17 free parameters and a total $\chi^2$/dof of 1.8.  }
\end{table}

In Table \ref{tab:parexpQCD1} we list the resonance parameters
for the fit in the explicit pQCD approach.
\begin{table}[htb]
\begin{tabular}{|c|c|c|c|}
\hline
Resonance & Mass [GeV] & $a_1$ [GeV$^{2}$] & $a_2$ [GeV$^{2}$] \\
\hline
\hline
$\omega$ & 0.782 & 0.616384 & 0.114681 \\
\hline
$\phi$ & 1.019 & 0.159562 & $-0.329255$ \\
\hline
$s_1$ & 1.799639 & 0.128654 & 0.026174 \\
\hline
\hline
$v_1$ & 1.000000 & $-0.309199$ & $-1.078695$ \\
\hline
$v_2$ & 1.627379 & 3.695960 & $-4.301057$ \\
\hline
$v_3$ & 1.779245 & $-3.693109$ & 3.630255 \\
\hline
\end{tabular}
\caption{\label{tab:parexpQCD1}Resonance parameters for the fit 
with explicit pQCD continuum. This fit has 14 free parameters and a
total $\chi^2$/dof of 2.0.}
\end{table}
The parameters for the explicit pQCD term are given in Table 
\ref{tab:parexpQCD2}.
\begin{table}[htb]
\begin{tabular}{|c|c|c|c|}
\hline
$a_{1}^{s}$ & $a_{1}^{v}$ & $b_{1}$ [GeV$^{-2}$] & $c_{1}$ [GeV$^{-1}$] \\
\hline
0.002321 & $-0.028391$ & 0.152903 & 0.161871 \\
\hline\hline
$a_{2}^{s}$ & $a_{2}^{v}$ & $b_{2}$ [GeV$^{-3}$] & $c_{2}$ [GeV$^{-1}$] \\
\hline
$-0.126598$ & $-0.011693$ & 1.159998 & 1.150000\\
\hline
\end{tabular}
\caption{\label{tab:parexpQCD2} Parameters of the explicit pQCD term 
for the fit with explicit pQCD continuum. This fit has 14 free parameters and a
total $\chi^2$/dof of 2.0.}
\end{table}

In Table \ref{tab:parexpCLAS1} we list the resonance parameters
for the fit including the preliminary data for $G_M^n$ from the
CLAS collaboration at Jefferson Lab \cite{CLASdata}
in the range $0.6$~GeV$^2$ $\simlt Q^2 \simlt 5$~GeV$^2$.
\begin{table}[htb]
\begin{tabular}{|c|c|c|c|}
\hline
Resonance & Mass [GeV] & $a_1$ [GeV$^{2}$] & $a_2$ [GeV$^{2}$] \\
\hline
\hline
$\omega$ & 0.782 & 0.669166 &  $-0.135957$\\
\hline
$s_1$ & 1.045277 & $-0.025807$ & 0.001144 \\
\hline
$s_2$ & 1.400423 & 0.261240 &  $-0.053588$\\
\hline
\hline
$v_1$ & 1.022008 & $-0.279441$ & $-1.215307$  \\
\hline
$v_2$ & 1.644552 & 3.823047 & $-4.561225$\\
\hline
$v_3$ & 1.770845 & $-3.849954$ & 4.027035 \\
\hline
\end{tabular}
\caption{\label{tab:parexpCLAS1}Resonance parameters for the fit 
including the preliminary CLAS data for  $G_M^n$
\cite{CLASdata}. This fit has 15 free parameters and a
total $\chi^2$/dof of 2.2.}
\end{table}
The parameters for the explicit pQCD term for the fit 
including the preliminary CLAS data for  $G_M^n$
are given in Table \ref{tab:parexpCLAS2}.

\begin{table}[htb]
\begin{tabular}{|c|c|c|c|}
\hline
$a_{1}^{s}$ & $a_{1}^{v}$ & $b_{1}$ [GeV$^{-2}$] & $c_{1}$ [GeV$^{-1}$] \\
\hline
$-0.000186$ & $-0.026941$ & 0.219241 & 0.169695 \\
\hline\hline
$a_{2}^{s}$ & $a_{2}^{v}$ & $b_{2}$ [GeV$^{-3}$] & $c_{2}$ [GeV$^{-1}$] \\
\hline
0.000527 & $-0.001835$ & 0.004155 & 0.106343\\
\hline
\end{tabular}
\caption{\label{tab:parexpCLAS2} Parameters of the explicit pQCD term 
for the fit including the preliminary CLAS data \cite{CLASdata}
for  $G_M^n$. 
This fit has 15 free parameters and a total $\chi^2$/dof of 2.2.}
\end{table}

In Table \ref{tab:parFW1}, we list the resonance parameters
for the fit with the bump-dip structure and additional low-mass 
strength allowed.
\begin{table}[htb]
\begin{tabular}{|c|c|c|c|c|}
\hline
Resonance & Mass [GeV] & $a_1$ [GeV$^{2}$] & $a_2$ [GeV$^{2}$] & 
$\Gamma$ [GeV]\\
\hline
\hline
$\omega$ & 0.782 & $-3.088199$ & 1.516336 & $-$\\
\hline
$s_1$ & 1.087524 & $-9.309347$  & 5.152311 & $-$\\
\hline
$s_2$ & 0.857075 & 7.969599 & $-3.102716$ & $-$\\
\hline
$s_3$ & 1.145783 & 5.332548 & $-3.754331$ & $-$\\
\hline
\hline
$v_1$ & 0.315028 & 0.002785 & $-0.008642$ & $-$\\
\hline
$v_2$ & 1.523890 & $-3.257202$ & 3.767630 & $-$\\
\hline
$v_3$ & 1.323997 & 2.770486 & $-5.497436$ & $-$\\
\hline
$v_4$ & 2.834388 & 0.177584 & $-0.011050$ & 13.477161\\
\hline
\end{tabular}
\caption{\label{tab:parFW1}
Resonance parameters for the fit with the bump-dip structure. 
This fit has 22 free parameters and a
total $\chi^2$/dof of 0.9 (space-like data only). Note that
$s_2$ and $v_1$ are the additional low-mass poles necessary
to generate the bump-dip structure.
}
\end{table}
The parameters for the explicit pQCD term in this fit
are given in Table \ref{tab:parFW2}.
\begin{table}[htb]
\begin{tabular}{|l|l|l|l|}
\hline
$a_{1}^{s}$ & $a_{1}^{v}$ & $b_{1}$ [GeV$^{-2}$] & $c_{1}$ [GeV$^{-1}$] \\
\hline
$-0.786259$ & $-0.320320$ & 0.971368 & 1.235451 \\
\hline\hline
$a_{2}^{s}$ & $a_{2}^{v}$ & $b_{2}$ [GeV$^{-3}$] & $c_{2}$ [GeV$^{-1}$] \\
\hline
$-0.000484$ & 0.033410 & 0.091209 & 0.994702\\
\hline
\end{tabular}
\caption{\label{tab:parFW2}
Parameters of the explicit pQCD term 
for the fit with bump-dip structure. This fit has 22 free parameters and a
total $\chi^2$/dof of 0.9 (space-like data only).}
\end{table}

\end{document}